\documentclass[pre,twocolumn,showpacs,groupedaddress,repreint]{revtex4-1}

\usepackage{graphicx}
\usepackage{dcolumn}
\usepackage{bm}
\usepackage{amsmath}
\usepackage{hyperref}
\usepackage{color}
\makeatother
\usepackage{tikz}
\begin{document}

\title{Transport and diffusion properties of Brownian particles powered by a rotating wheel}

\author{Bao-quan  Ai} \email[Email: ]{aibq@scnu.edu.cn}
\affiliation{Guangdong Provincial Key Laboratory of Quantum Engineering and Quantum Materials, School of Physics and Telecommunication
Engineering, South China Normal University, Guangzhou 510006, China.}

\date{\today}

\begin{abstract}
   \indent Diffusion and rectification of Brownian particles powered by a rotating wheel are numerically investigated in a two-dimensional channel. The nonequilibrium driving comes from the rotating wheel, which can break thermodynamical equilibrium and induce the directed transport in an asymmetric potential. It is found that the direction of the transport along the potential is determined by the asymmetry of the potential and the position of the wheel. The average velocity is a peaked function of the angular speed (or the diffusion coefficient) and the position of the peak shifts to large angular speed(or diffusion coefficient) when the diffusion coefficient (or the angular speed) increases. There exists an optimal angular speed (or diffusion coefficient) at which the effective diffusion coefficient takes its maximal value. Remarkably, the giant acceleration of diffusion is observed by suitably adjusting the system parameters. The parameters corresponding to the maximum effective diffusion coefficient are not the same as the parameters at which average velocity is maximum.
\end{abstract}
\pacs{05. 40. -a, 82. 70. Dd}
\maketitle

\section{Introduction}
\indent  Thermal noise induced by random collision of the liquid molecules is usually an obstacle for controlling particle flow at nanoscopic scales. However, in a ratchet system, Brownian particles can move in a periodic structure with nonzero macroscopic velocity  in the absence of macroscopic force on average\cite{hanggi,Reimann}. Experimental realizations of ratchets are spread over many different fields of physics \cite{Salger,Roche}, chemistry \cite{Wilson,Astumian0}, and biology \cite{Sokolova} where the nonequilibrium fluctuations may be intrinsic to it (e.g., nonthermal noise), or external to the system (e.g., induced by an experimentalist). Besides fundamental interest in biology because they play
a role in biological protein motors\cite{Collins}, these devices have also attracted great interest as the basic working principle of practical devices. Therefore, the implications of rectification has attracted the interest of researchers in a variety of fields, ranging from biology to nanoscience, due to its relevance in the transport and motion at small scales.

\indent Usually, the ratchet setup requires three critical ingredients \cite{Denisov}: (1)asymmetry (spatial and/or temporal), which can violate the left-right symmetry of the response), (2)nonlinearity, it is necessary since the system will produce a zero mean output from zero-mean input in a linear system, and (3)a zero-mean nonequilibrium driving, it should break thermodynamical equilibrium, which forbids appearance of a directed transport due to the Second Law of Thermodynamics. Depending on the type of the nonequilibrium driving, ratchet devices fall into four categories. (a)Tilting ratchets, where the nonequilibrium driving is either an unbiased periodic force or an unbiased stationary random process, such as rocking ratchets\cite{Magnasco,Grossert}, fluctuating force ratchets\cite{Bartussek} and so on. (b)Pulsating ratchets, where the substrate potential was modulated by the time-dependent periodic process, such as on-off ratchets\cite{Bug,Lau}, fluctuating potential ratchets\cite{Astumian}, traveling potential ratchets\cite{Borromeo,Sandor} and so on. (c)Coupled ratchets \cite{Derenyi,Silva,Zheng}, where collective effects arise when several copies of single ratchet system interacts with each other. (d)Self-propelled ratchets\cite{Reichhardt,Kummel,Ghosh,Wan,Ai1,Ai2,Ai3,Kaiser}, where active particles interact with an asymmetric substrate, a net directed motion can arise even without external driving, for example active swimmers can be rectified in a two-dimensional periodically compartmentalized channel\cite{Ghosh}. In these ratchet systems, the nonequilibrium driving can break thermodynamical equilibrium and induce the directed transport.

\indent In most ratchet systems, the nonequilibrium driving directly acts on each particle. However, in some situations, the nonequilibrium driving indirectly acts on each particle (such as the temperature gradient driving and the driving from local turbulent medium), where the nonequilibrium driving is transferred from the driving source to each particle through the interactions between particles.  Therefore, it would be interesting to study the nonequilibrium transport of the interacting particles in the presence of a local driving source. In this paper, we propose a different ratchet mechanism, where the nonequilibrium driving comes from the rotating wheel. The driving force was transferred from the rotating wheel to Brownian particles, which can break thermodynamical equilibrium and induce the directed transport in the asymmetric potential. We focus on finding how the system parameters affect the rectification and diffusion of Brownian particles.

\section{Model and methods}
\begin{figure}[htpb]
\vspace{1cm}
  \label{fig1}\includegraphics[width=0.8\columnwidth]{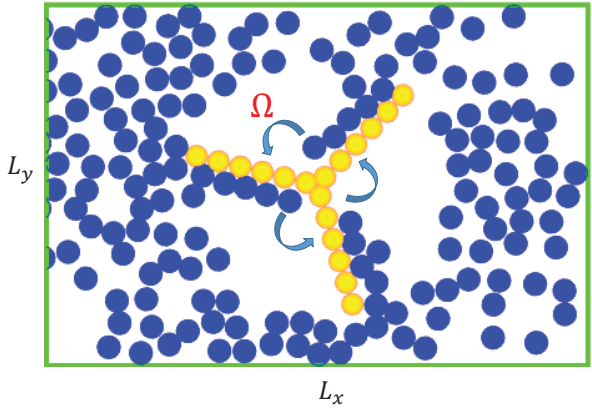}
  \caption{(a) (Color online) Schematic of the rotating wheel motor: Brownian particles (blue disks) driven by the rotating wheel with angular speed $\Omega$ and moving in a straight periodic channel with width $L_y$ and period $L_x$. Periodic boundary conditions imposed in the $x$-direction and hard wall boundary conditions in the $y$-direction. The wheel is composed of three paddles and each paddle consists of $n_p$ paddle particles (yellow disks). The wheel is centered at the point $(x_c,y_c)$ and the angle between paddles is $2\pi/3$. }
\end{figure}

\indent We consider $N$ Brownian particles with radius $r$ moving in a two-dimensional straight channel with period $L_x$ and width $L_y$  as showed in Fig. 1.  In the channel, the confining walls are located at $y = 0$ and $y = L_y$ and particles experience a substrate potential $V(x)=V(x+L_x)$ along $x$-direction. In addition, Brownian particles are driven by collisions with the paddles of the wheel. The paddles rotate around the point $(x_c,y_c)$ with angular speed $\Omega$. The wheel is composed of three paddles and each paddle consists of $n_p$  paddle particles with radius $R$. The forces acting particle $i$ from the other particles and from the paddle particles are respectively defined as $\mathbf{F}_i=F^{x}_{i}\mathbf{e}_x+F^{y}_{i}\mathbf{e}_y=\sum_{j}\mathbf{F}_{ij}$ and  $\mathbf{G}_{i}=G^{x}_{i}\mathbf{e}_x+G^{y}_{i}\mathbf{e}_y=\sum_{j}\mathbf{G}_{ij}$.  The dynamics of particle $i$ is described by the following overdamped Langevin equations

\begin{equation}\label{e1}
\frac{dx_i}{dt}=\mu[F^x_i+G^x_i+f_p(x_i)]+\sqrt{2D_0}\xi^x_i(t),
\end{equation}

\begin{equation}\label{e2}
\frac{dy_i}{dt}=\mu[F^y_i+G^y_i]+\sqrt{2D_0}\xi^y_i(t),
\end{equation}
where $\mu$ is the mobility and $D_0$ denotes the  diffusion coefficient. Thermal fluctuations due to the coupling of the particle with the environment are modeled by zero-mean, Gaussian white noise $\xi_i(t)$ with autocorrelation function $\langle \xi^\alpha_{i}(t)\xi^\beta_{j}(s)\rangle = \delta_{\alpha\beta}\delta_{ij}\delta(t-s)$, where $\alpha,\beta=x, y$. $\delta$ is the Dirac delta function and the symbol $\langle...\rangle$
denotes an ensemble average over the distribution of the random forces.

\indent The particle-particle interaction and particle-paddle interaction are assumed to be of the linear spring form with the stiffness constant $k_1$ and $k_2$, respectively. $\mathbf{F}_{ij}=k_1(2r-r_{ij})\mathbf{e}_{r}$, if $r_{ij}<2r$ ($\mathbf{F}_{ij}=0$ otherwise), and $r_{ij}$ is the distance between Brownian particle $i$ and $j$. $\mathbf{G}_{ij}=k_2[(r+R)-r_{ij}]\mathbf{e}_{r}$, if $r_{ij}<r+R$ ($\mathbf{G}_{ij}=0$ otherwise), and $r_{ij}$ is the distance between Brownian particle $i$ and paddle particle $j$. In order to mimic hard particles, we use large values of $k_1$ and $k_2$, thus ensuring that particle overlaps decay quickly.

\begin{figure}[htpb]
\vspace{1cm}
  \label{fig1}\includegraphics[width=0.7\columnwidth]{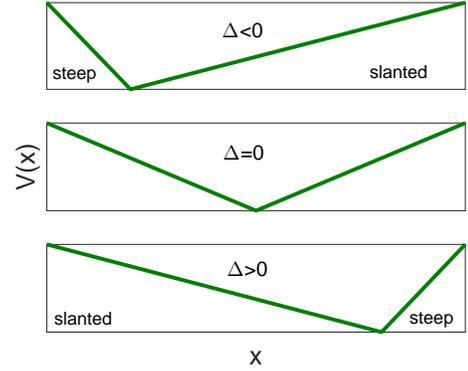}
  \caption{(Color online) The profile of $x$-direction the potential $V(x)$ described in Eq. (\ref{Vx}) for different values of the parameter $\Delta$. The left side of the potential is steep for $\Delta<0$ and the right side is steep for $\Delta>0$. }
\end{figure}

\indent $f_p(x)$ is the substrate force along $x$-direction which arises from a periodic potential $V(x)$ (shown in Fig. 2). The profile of the potential within one period is described by
\begin{equation}\label{Vx}
     V(x)=
\begin{cases}
\frac{V_0}{L_1}(L_1-x), & \text{$0\leq x<L_1$};\\
\frac{V_0}{L_2}(x-L_1), & \text{$L_1\leq x\leq L_x$},
\end{cases}
\end{equation}
where $V_0$ is the height of the potential. The asymmetry of the potential is defined as $\Delta=(L_1-L_2)/L_x$. The left side of the potential is steep for $\Delta<0$ and the right side is steep for $\Delta>0$.

\indent By introducing the characteristic length  and time scales: $\hat{x}=\frac{x}{2r}$, $\hat{y}=\frac{y}{2r}$, $\hat{t}=\frac{t}{\tau_0}$, and $\tau_0=\frac{4r^2}{\mu V_0}$, Eqs.(\ref{e1},\ref{e2}) can be rewritten in the dimensionless forms
\begin{equation}\label{e3}
 \frac{d\hat{x}_i}{d\hat{t}}=\hat{F}^{\hat{x}}_i+\hat{G}^{\hat{x}}_i+\hat{f}_p(\hat{x}_i)+\sqrt{2\hat{D}_0}\hat{\xi}^{\hat{x}}_i(\hat{t}),
\end{equation}

\begin{equation}\label{e4}
 \frac{d\hat{y}_i}{d\hat{t}}=\hat{F}^{\hat{y}}_i+\hat{G}^{\hat{y}}_i+\sqrt{2\hat{D}_0}\hat{\xi}^{\hat{y}}_i(\hat{t}),
\end{equation}
and the other parameters are $\hat{L}_x=\frac{L_x}{2r}$, $\hat{L}_y=\frac{L_y}{2r}$, $\hat{R}=\frac{R}{2r}$, $\hat{k}_1=\frac{4k_1r^2}{V_0}$, $\hat{k}_2=\frac{4k_2r^2}{V_0}$, $\hat{D}_0=\frac{D_0}{\mu V_0}$ and $\hat{\Omega}=\Omega\tau_0$. The periodic potential can be rewritten as $\hat{V}(\hat{x})=1-\hat{x}/{\hat{L}_1}$ for $0\leq \hat{x}<\hat{L}_1$ and $(\hat{x}-\hat{L}_1)/{\hat{L}_2}$ for $\hat{L}_1\leq \hat{x}\leq 1$, where $\hat{L}_1=\frac{L_1}{2r}$ and $\hat{L}_2=\frac{L_2}{2r}$. From now on, we will use only the dimensionless variables and shall omit the hat for all quantities occurring in Eqs. (\ref{e3},\ref{e4}).

\indent The behavior of the quantities of interest can be corroborated by integration of the Langevin Eqs.(\ref{e3},\ref{e4}) using the second-order stochastic Runge-Kutta algorithm. Because particles are confined in the $y$-direction, directed transport only occurs in the $x$-direction. In the asymptotic long-time regime, the average velocity of the particle along $x$-direction can be obtained from the following formula
\begin{equation}\label{velocty}
  V_{x}=\lim_{t\rightarrow\infty}\frac{\langle x_i(t)\rangle}{t},
\end{equation}
and the effective diffusion coefficient $D_x$ along $x$-direction can be calculated by the formula
\begin{equation}\label{DX}
  D_x=\lim_{t\rightarrow\infty}\frac{\langle[x_i(t)-\langle x_i(t)\rangle]^2\rangle}{2t},
\end{equation}
where the symbol $\langle...\rangle$ denotes an average over particle number $N$. We use the scaled effective diffusion coefficient $D_{eff}=D_x/D_0$ for convenience. In addition, we define the ratio between the area occupied by particles and the total available area as the packing fraction $\phi=\pi(3n_p R^2+N r^2)/(L_x L_y)$.

\section{Results and Discussion}
\indent In our simulations, we have considered more than $10^4$ realizations to improve accuracy and minimize statistical errors. The integration step time was chosen to be smaller than $10^{-4}$ and the total integration time was more than $10^7$.  Unless otherwise noted, our simulations are under the parameter sets: $L_x=24.0$, $L_y=16.0$, $x_c=L_x/2.0$, $y_c=L_y/2.0$, $k_1=100.0$, $k_2=500.0$, $R=0.5$, $N=60$, and $n_p=6$. We vary the parameters and measure the average velocity and the effective diffusion coefficient of Brownian particles.

\subsection{Longitudinal direction rectification and diffusion}
\begin{figure}[htpb]
\vspace{1cm}
  \label{fig2}\includegraphics[width=0.8\columnwidth]{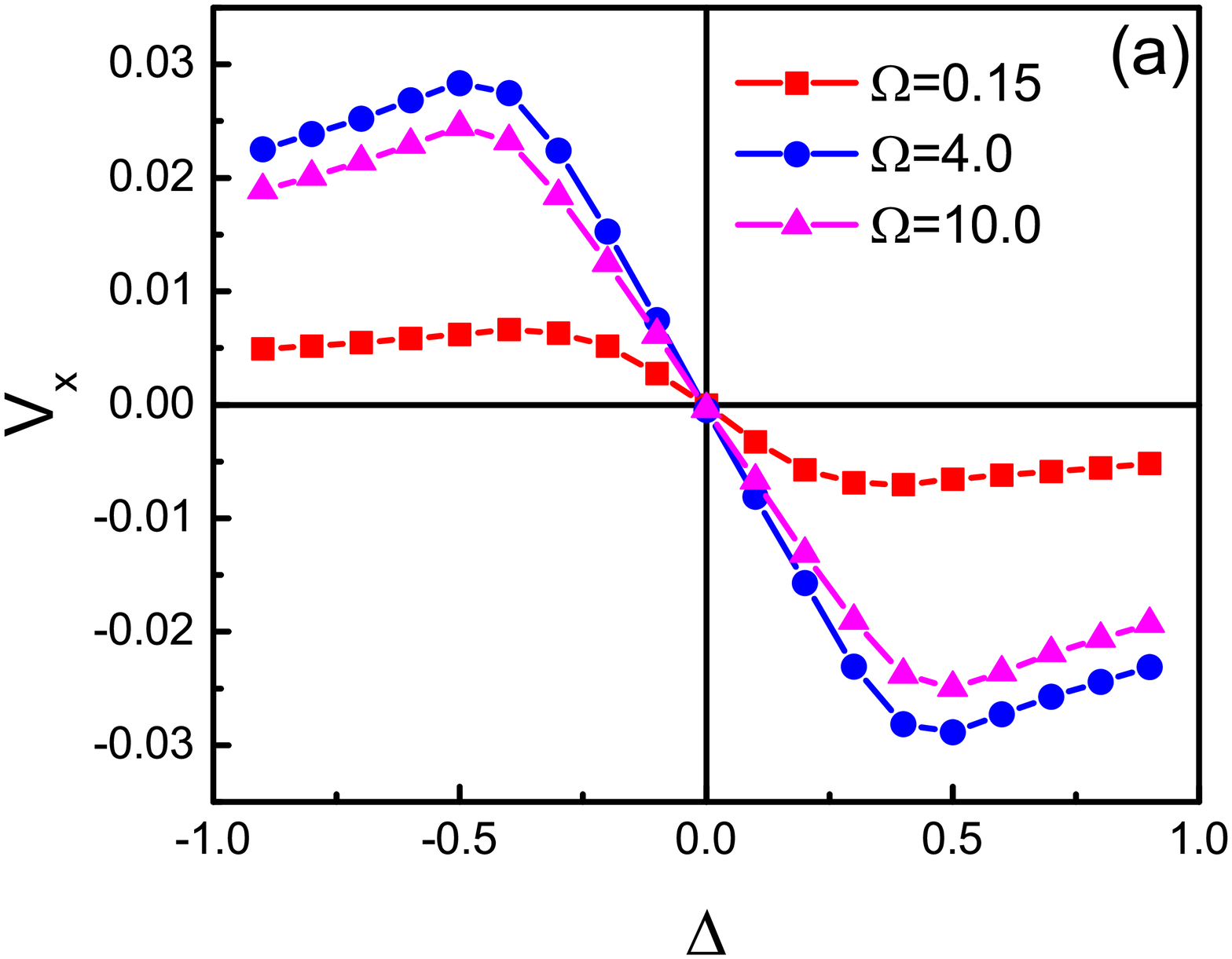}
  \includegraphics[width=0.8\columnwidth]{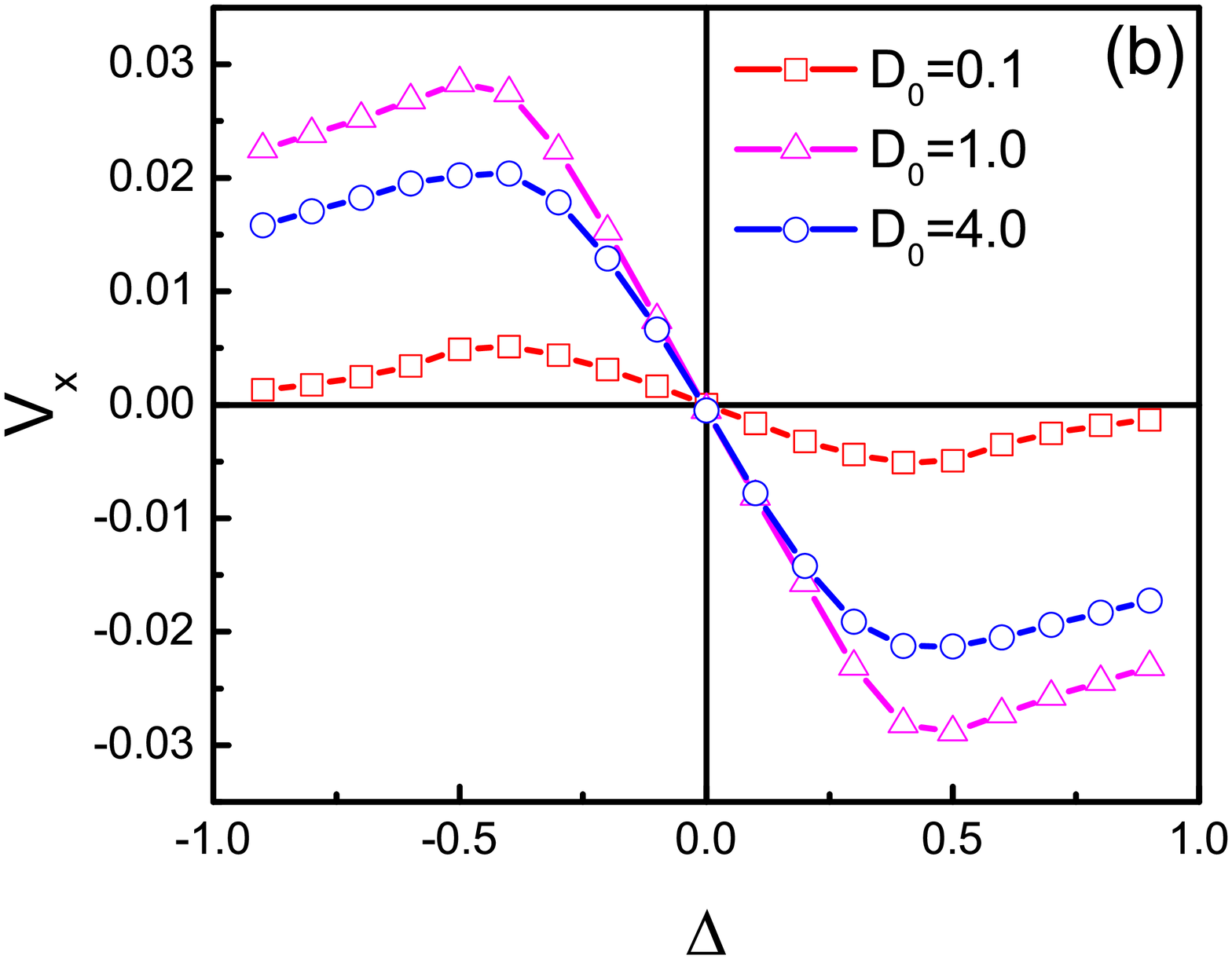}
  \caption{(Color online)Average velocity $V_s$ as a function of the asymmetric parameter $\Delta$. (a)For different values of $\Omega$ at $D_0=1.0$. (b)For different values of $D_0$ at $\Omega=4.0$.}
\end{figure}

\indent We firstly discuss the dependence of directed transport on the asymmetry of the potential for the case of $\Omega>0$. The average velocity $V_x$ as a function of the asymmetric parameter $\Delta$ is shown in Figs. 3(a) and 3(b). It is found that average velocity $V_x$ is positive for $\Delta<0$, zero at $\Delta=0$, and negative for $\Delta>0$. When the potential is completely symmetric (shown in the middle panel of Fig. 2), the probabilities of crossing right and left barriers are the same, so the average velocity on average is equal to zero. When $\Delta<0$, the left side from the minima of the potential is steeper (shown in the top panel of Fig. 2), it is easier for particles to move toward the slanted side than toward the steep side, so particles on average move to the right ($V_x>0$). Similarly, particles on average move to the left ($V_x<0$) when $\Delta>0$. Therefore, the asymmetry of the potential determines the direction of the transport.

\begin{figure}[htpb]
\vspace{1cm}
  \label{fig3}\includegraphics[width=0.8\columnwidth]{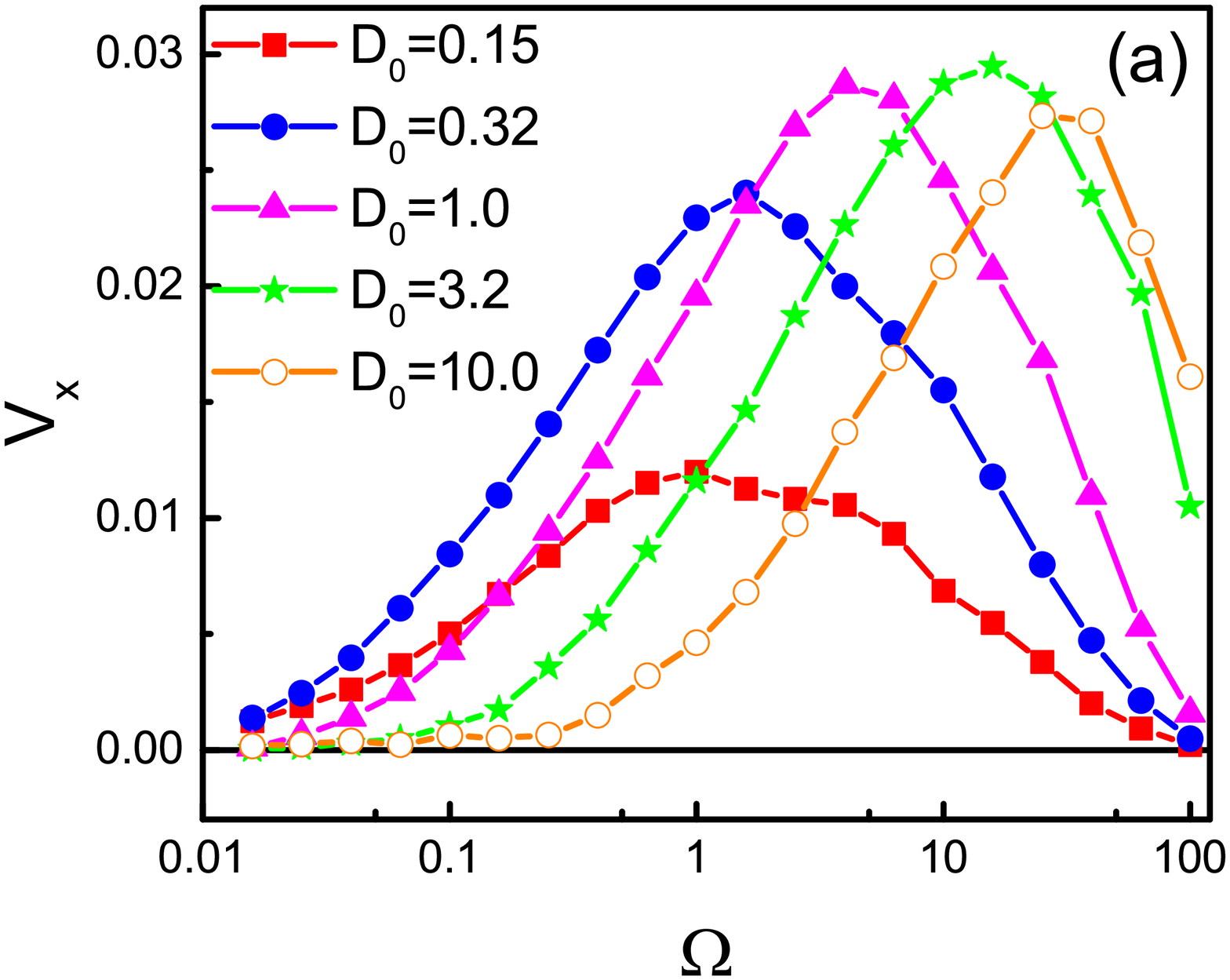}
  \includegraphics[width=0.8\columnwidth]{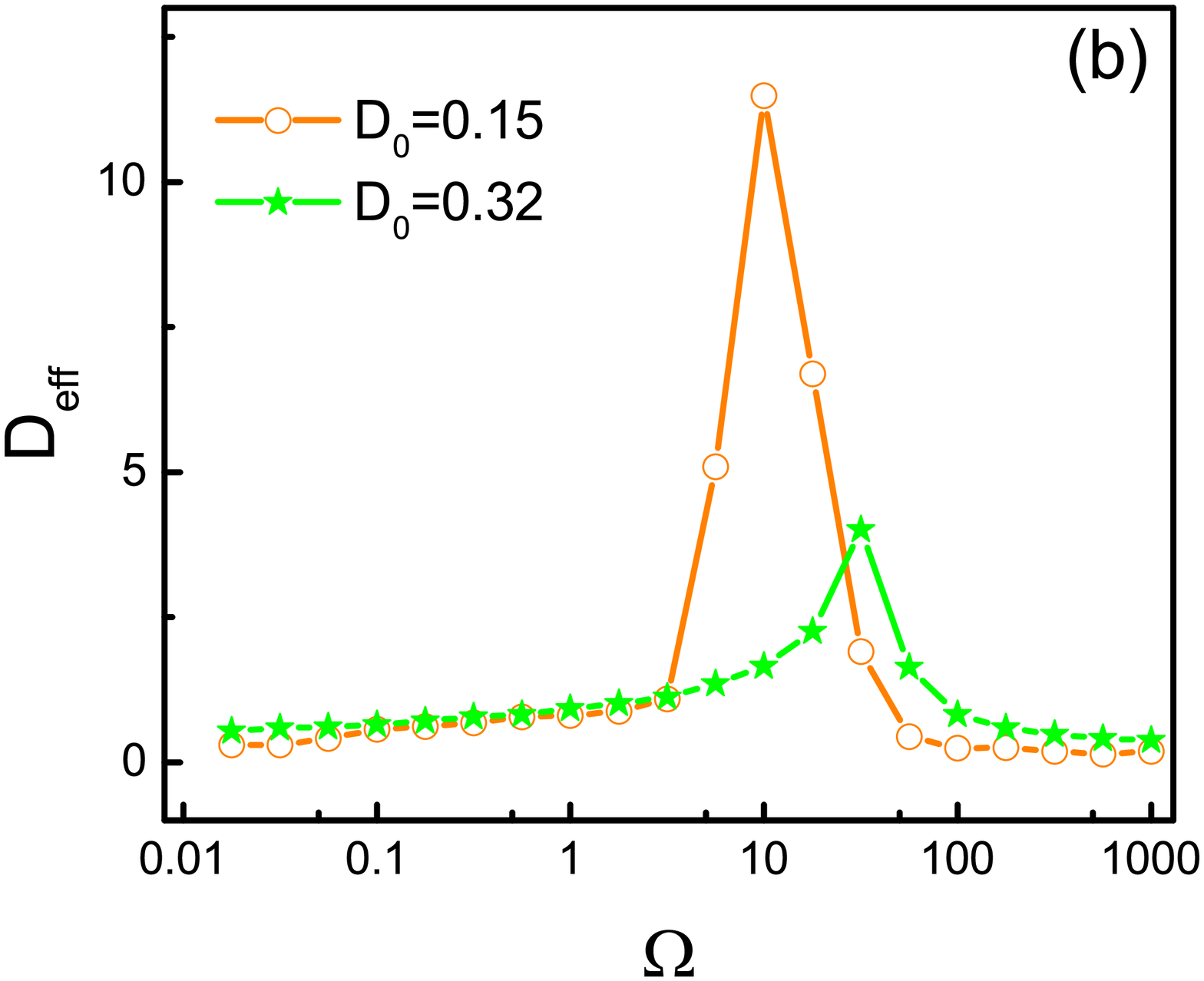}
  \includegraphics[width=0.8\columnwidth]{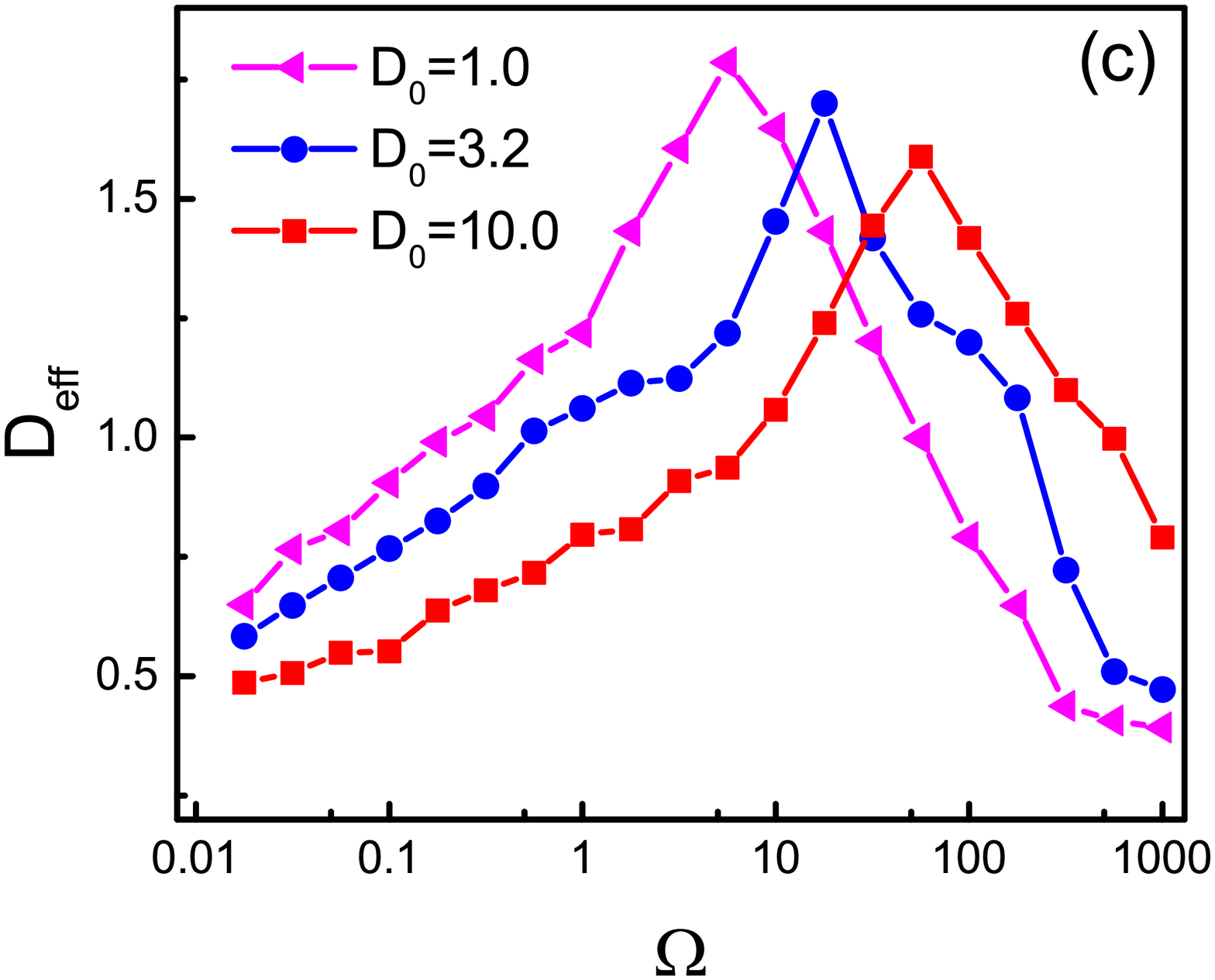}
  \caption{(a) (Color online)(a) Average velocity $V_x$ as a function of the angular speed $\Omega$ for different values of $D_0$ at $\Delta=-0.5$.  (b)and (c) Scaled effective diffusion coefficient $D_{eff}$ as a function of the angular speed $\Omega$ for different values of $D_0$ at $\Delta=-0.5$.}
\end{figure}

\indent Figure 4(a) shows the average velocity $V_x$ as a function of the angular speed $\Omega$ for different values of $D_0$. The curves are observed to be bell shaped, which show the feature of resonance. When $\Omega\rightarrow 0$, the paddles rotate very slowly, the nonequilibrium driving from the paddles can be neglected and the ratchet effect disappears, thus $V_x $ tends to zero.  When $\Omega\rightarrow \infty$, the paddles rotate very fast, the whole wheel can be regarded as a rotating hard disk and particles cannot enter the region of wheel, particles cannot obtain the driving force from the paddles, so the average velocity goes to zero. Therefore, there exists an optimal value of $\Omega$ at which the average velocity $V_x$ takes its maximal value. In addition, the position of the peak shifts to large $\Omega$ when $D_0$ increases.

\indent Figures 4(b) and 4(c) describe the scaled effective diffusion coefficient $D_{eff}$ a function of the angular speed $\Omega$ for different values of $D_0$. The most important feature is the peak near the critical angular speed $\Omega_c$, which gets more and more pronounced as $D_0$ decreases. Interestingly, when $D_0<1.0$ (shown in Fig. 4(b)) the giant acceleration of diffusion is observed at the critical $\Omega_c$, which is a universal phenomenon predicted by the theoretical analysis \cite{Costantini,Reimann2}. The basic physical mechanism behind this effect may be explained as follows. When $D_0$ is small ($D_0<1.0$, e.g. $0.15$), particles become trapped between potential barriers, the particle-paddle interaction dominates the diffusion. In this case, at the critical $\Omega_c$, particles can be thrown out from the minima of the potential and can diffuse quickly through the potential, which leads to a large value of $D_e$.  Therefore, the scaled effective diffusion coefficient $D_{eff}=D_e/D_0$ is much larger than $1$, which indicates the giant acceleration of diffusion. In addition, the angular speed corresponding to the maximum effective diffusion coefficient $D_{eff}$ is not the same as the angular speed at which average velocity $V_x$ is maximum. Note that in the previous works\cite{Costantini,Reimann2,Kim,Hayashi,Lindner} the giant acceleration of diffusion usually appears in a \emph{tilted} periodic potential, however, in the present system the appearance of the giant acceleration of diffusion cannot require a constant force.

\indent Note that the rotation of the wheel in the present system can create the turbulence and in the turbulent medium the effective coefficient of turbulent thermal diffusion\cite{Amir}can be obtained from the approximate method. Here, we use an alternative approximate method to obtain the the bulk diffusivity of Brownian particles driving by the rotating wheel. In the bulk case, particles in the present system can be approximately seen as active particles undergoing a constant speed  $v_0$ ($\propto \Omega$) and additionally driven by a random torque with the mean value $\Omega$. Following the approach in Ref. \cite{Ao}, we can approximately obtain the bulk diffusivity $D_{bulk}=D_0+\frac{aD_0\Omega^4}{D_0^2+b\Omega^6}$, where the positive variables $a$ and $b$ depend on the other parameters of the system. The bulk diffusivity $D_{bulk}$ tends to $D_0$ when $\Omega\rightarrow 0$ or $\Omega\rightarrow \infty$, thus the bulk diffusivity $D_{bulk}$ of Brownian particles is a peaked function of the angular speed $\Omega$ which can be confirmed by the numerical simulations.

\begin{figure}[htpb]
\vspace{1cm}
  \label{fig3}\includegraphics[width=0.8\columnwidth]{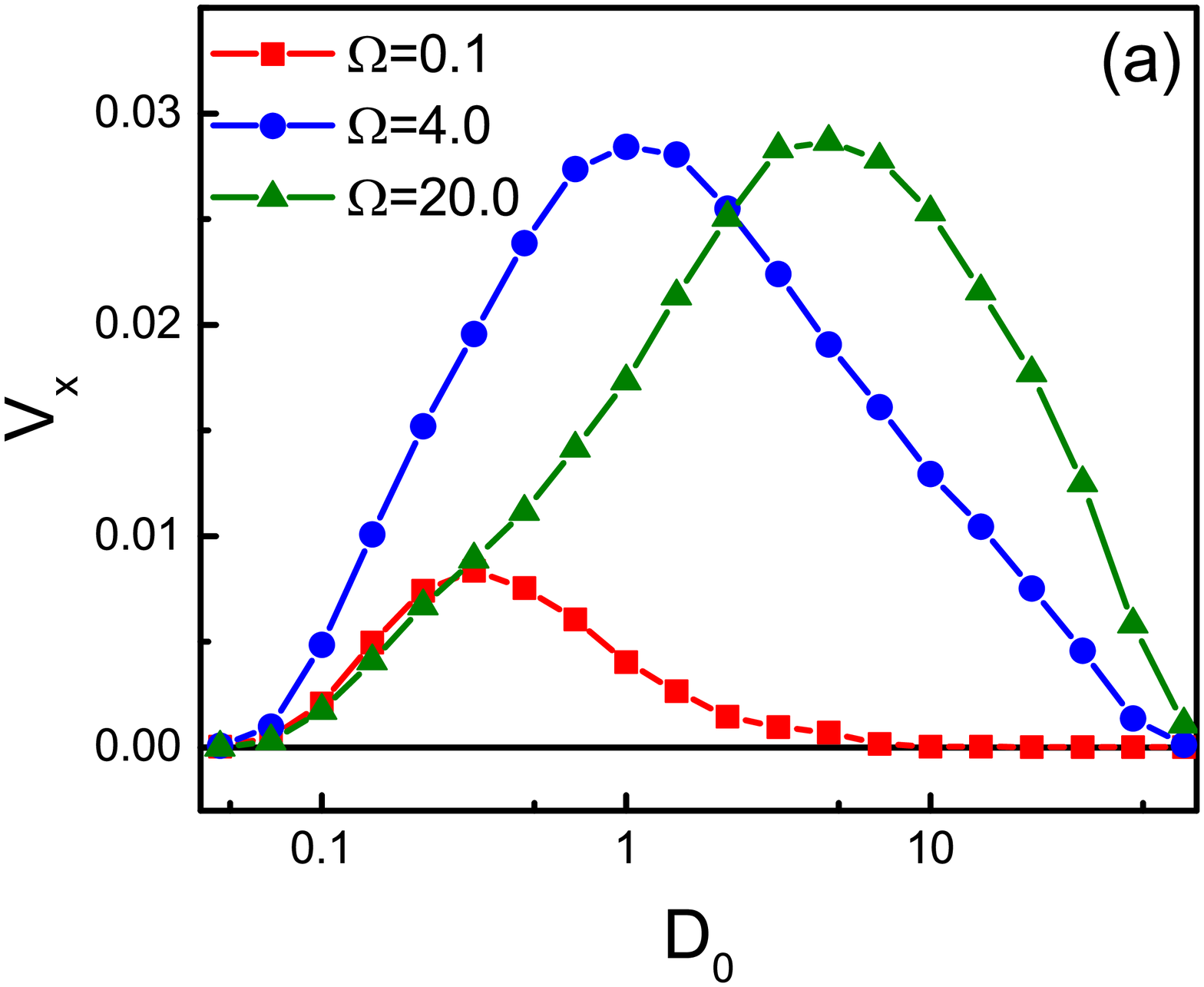}
  \includegraphics[width=0.8\columnwidth]{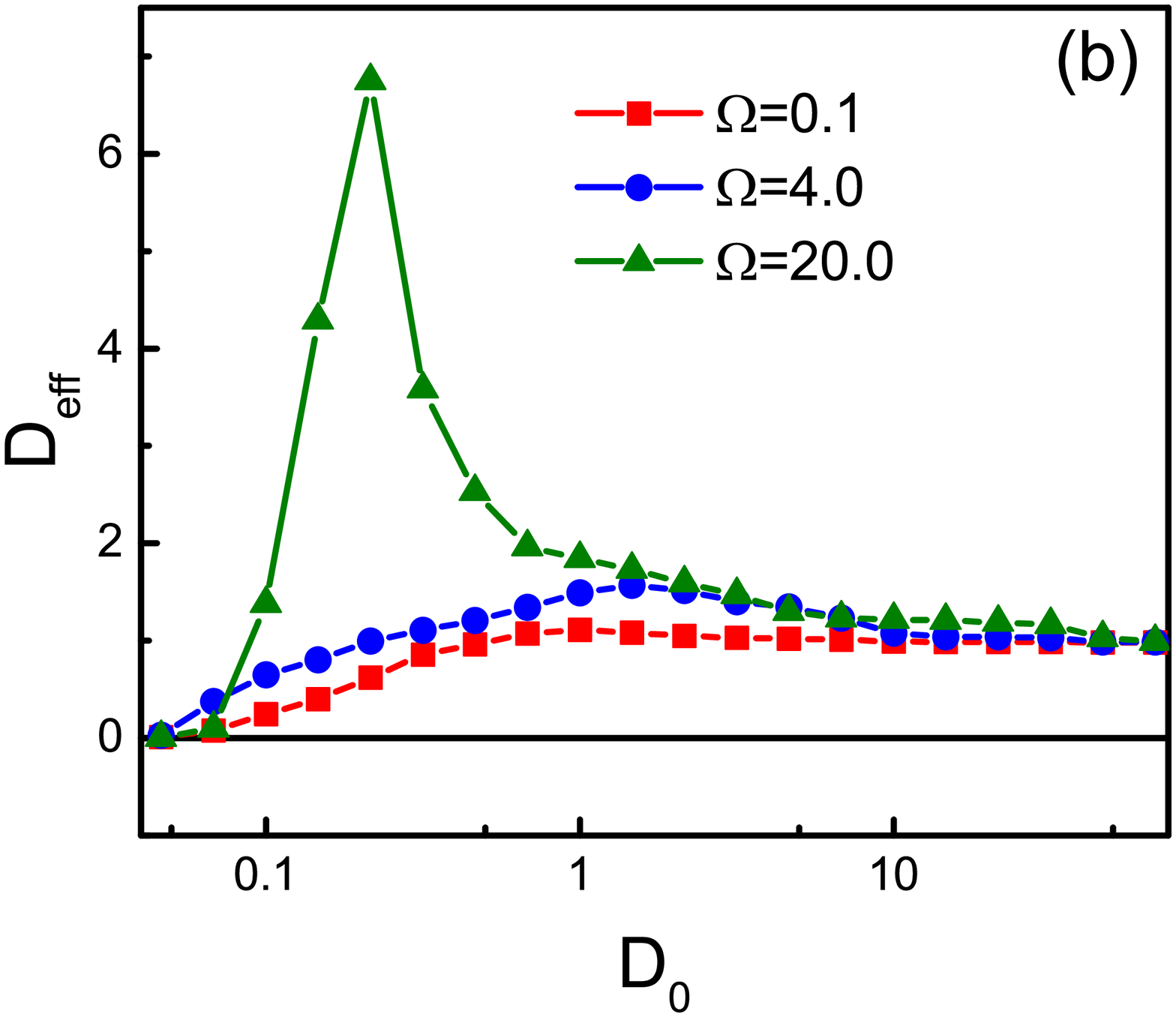}
  \caption{(a) (Color online)(a) Average velocity $V_x$ as a function of the diffusion coefficient $D_0$ for different values of $\Omega$ at $\Delta=-0.5$. (b) Scaled effective diffusion coefficient $D_{eff}$ as a function of the diffusion coefficient $D_0$ for different values of $\Omega$ at $\Delta=-0.5$.}
\end{figure}

\indent Figure 5(a) shows the average velocity $V_x$ as a function of the diffusion coefficient $D_0$ for different values of $\Omega$. It is found that average velocity is a peaked function of the diffusion coefficient. When $D_0\rightarrow 0$, particle cannot pass across the barrier of the potential, so there is no net current. When $D_0\rightarrow \infty$, the potential can be neglected and the ratchet effect disappears, so the average velocity goes to zero. Therefore, there exists an optimal value of $D_0$ at which the average velocity $V_x$ is maximal. Similar to Fig. 4(a), the position of the peak shifts to large $D_0$ when $\Omega$ increases.

\indent Figure 5(b) describes the scaled effective diffusion coefficient $D_{eff}$ a function of the diffusion coefficient $D_0$ for different values of $\Omega$. It is noted that the scaled effective diffusion coefficient as a function of $D_0$ exhibits a pronounced resonance peak in the curves. When $D_0\rightarrow 0$, particles are trapped in the valley of the potential, so  $D_{eff}$ tends to zero. When $D_0\rightarrow \infty$, the potential can be neglected and particles diffuse freely, $D_x$ tends to $D_0$, so $D_{eff}\rightarrow 1$. There exists a finite value of $D_0$ at which $D_{eff}$ takes its maximal value. The most remarkable feature is the huge enhancement of diffusion at $\Omega=20.0$, where the scaled effective diffusion coefficient is equal to $6.7$ at $D_0=0.2$. Note that $D_0$ corresponding to the maximum effective diffusion coefficient is not the same as $D_0$ at which average velocity is maximum, which is similar to the mismatch between optimal rectification speed and diffusion in Ref. \cite{Machura}.

\begin{figure}[htpb]
\vspace{1cm}
  \label{fig3}\includegraphics[width=0.8\columnwidth]{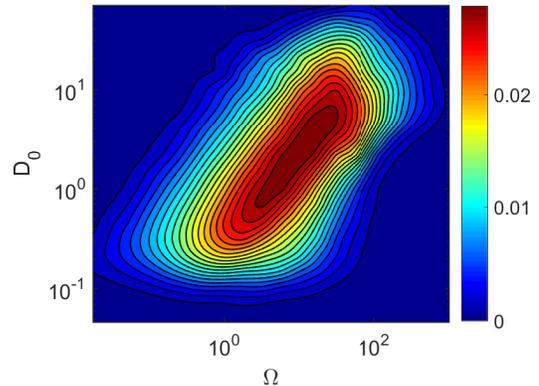}
  \caption{(Color online)Contour plots of the average velocity $V_x$ as a function of the system parameters $\Omega$ and $D_0$ at $\Delta=-0.5$.}
\end{figure}

\indent To study in more detail the dependence of the average velocity on $\Omega$ and $D_0$, we plotted contour plots of average velocity $V_x$ as a function of the system parameters $\Omega$ and $D_0$ in Fig. 6.
It is found that the average velocity tends to zero when $D_0$ ($\Omega$)$\rightarrow 0$ or $\infty$. There exist finite values of $D_0$ and $\Omega$ where average velocity takes its maximal value. Moreover, the position of the peak shifts to large $D_0$ (or $\Omega$) when $\Omega$ (or $D_0$) increases. Therefore, the optimal $D_0$ and $\Omega$ can facilitate rectified transport of Brownian particles.

\begin{figure}[htpb]
\vspace{1cm}
  \label{fig3}\includegraphics[width=0.8\columnwidth]{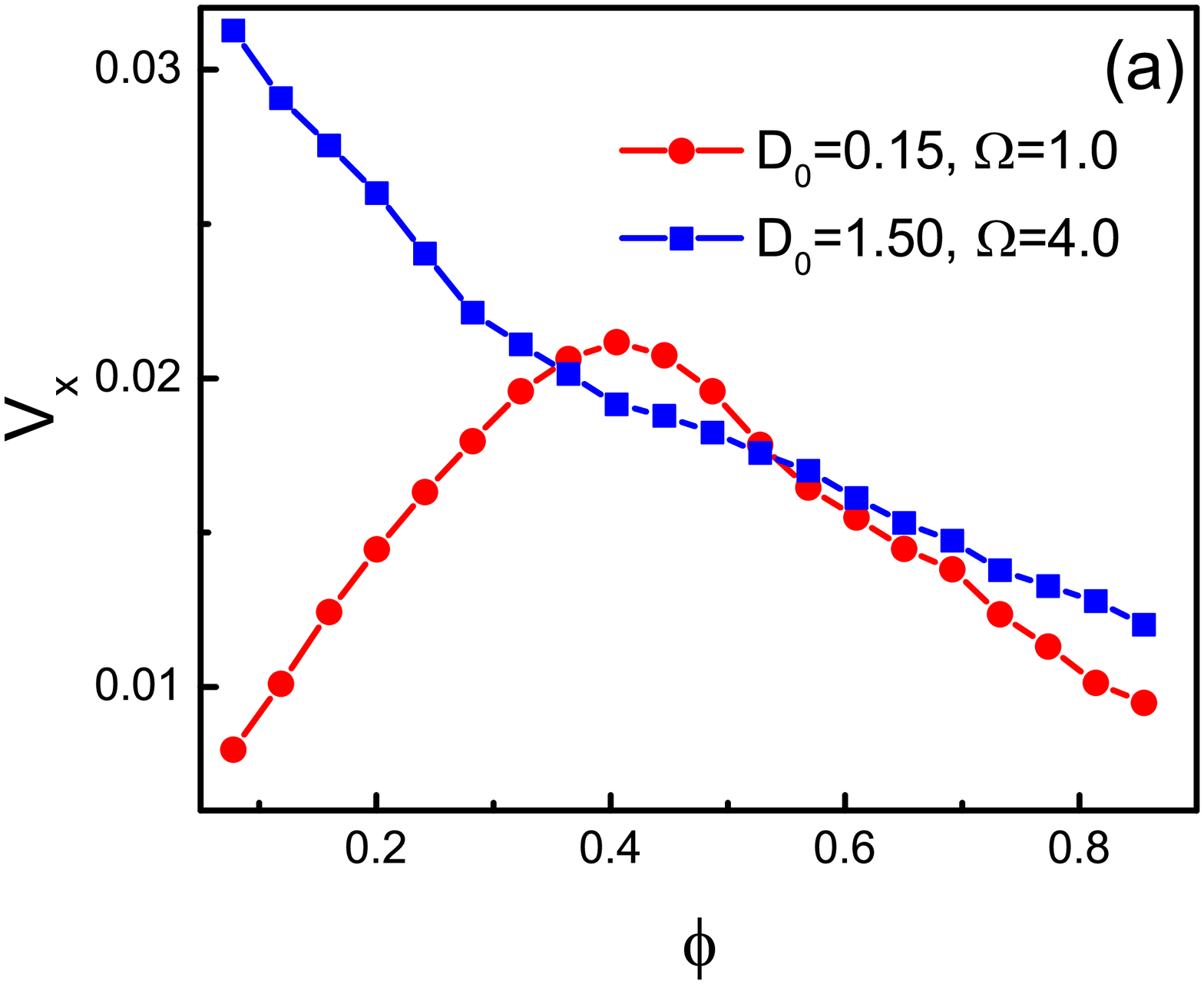}
  \includegraphics[width=0.8\columnwidth]{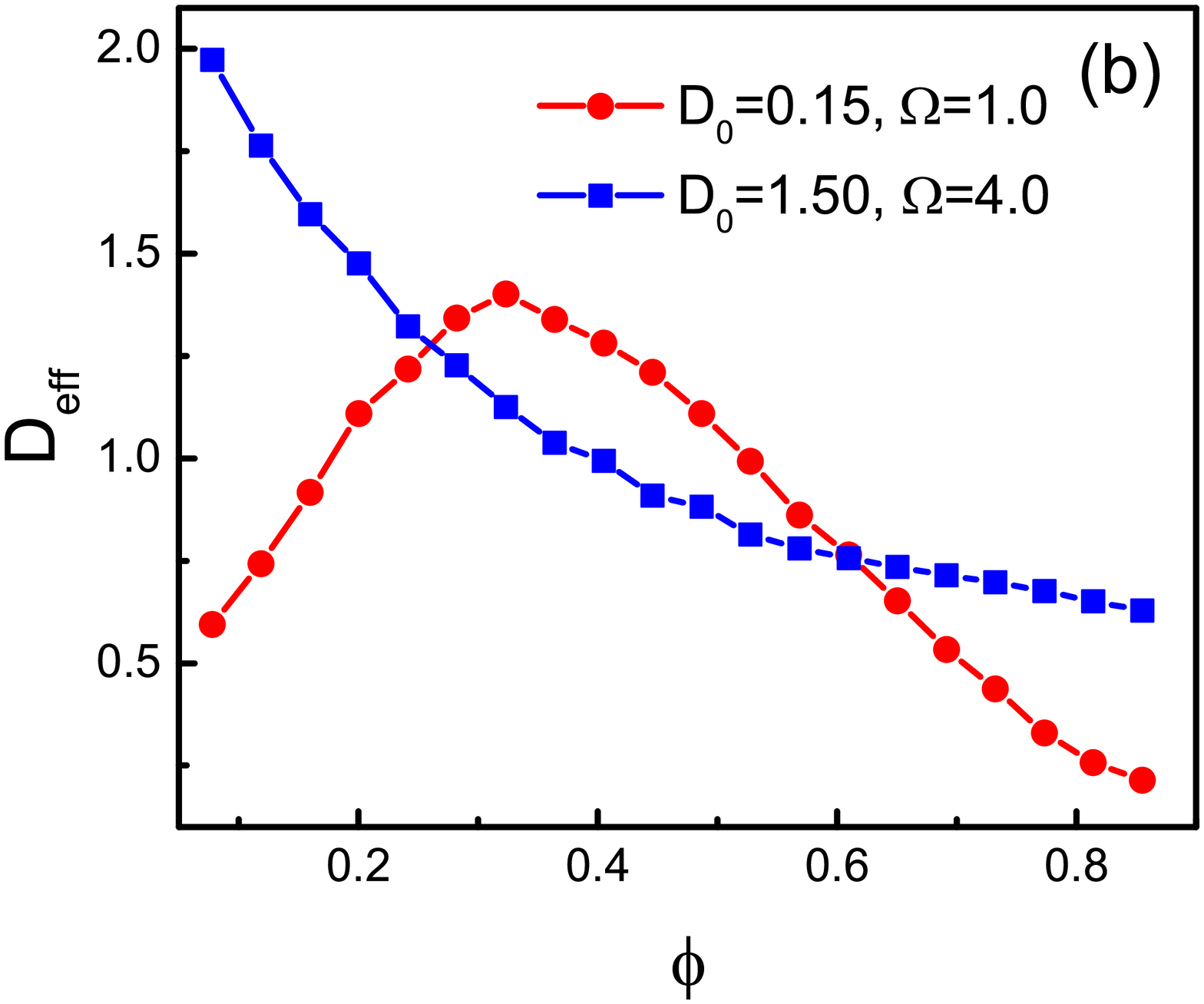}
  \caption{(Color online)(a)Average velocity $V_x$ as a function of the packing fraction $\phi$ at $\Delta=-0.5$. (b)Scaled effective diffusion coefficient $D_{eff}$ as a function the packing fraction $\phi$ at $\Delta=-0.5$.}
\end{figure}

\indent Figure 7(a) describes the dependence of the average velocity on the packing fraction for two different cases:(I)particles can easily pass across the barrier ($D_0=1.50$ and $\Omega=4.0$) and (II) particles cannot easily pass across the barrier($D_0=0.15$ and $\Omega=1.0$). An increase of the packing fraction can cause two results: (A)activating motion in an analogy with the thermal noise activated motion for a single stochastically driven ratchet, which facilitates the ratchet transport and (B)reducing the mobility of particles, which blocks the ratchet transport. For case I ($D_0=1.50$ and $\Omega=4.0$), particles can easily pass across the barrier, the factor B always dominates the transport, thus the average velocity $V_x$ decreases monotonically with increasing $\phi$. For case II ($D_0=0.15$ and $\Omega=1.0$), when the packing fraction increases from zero, the factor A first dominates the transport, so average velocity increases with the packing fraction. However, when the packing fraction become large, the factor B dominates the transport, thus average velocity decreases with increasing $\phi$. Therefore, for case II, there exists an optimal value of $\phi$ at which the average velocity takes its maximal value. Similar to Fig. 7 (a), the effective diffusion coefficient decreases with increasing $\phi$ for case I and the effective diffusion coefficient is a peaked function of $\phi$ for case II (shown in Fig. 7 (b)).

\subsection{Dependence of rectification on $L_y$, $k_1$, and $x_c$}

\begin{figure}[htpb]
\vspace{1cm}
  \label{fig8}\includegraphics[width=0.8\columnwidth]{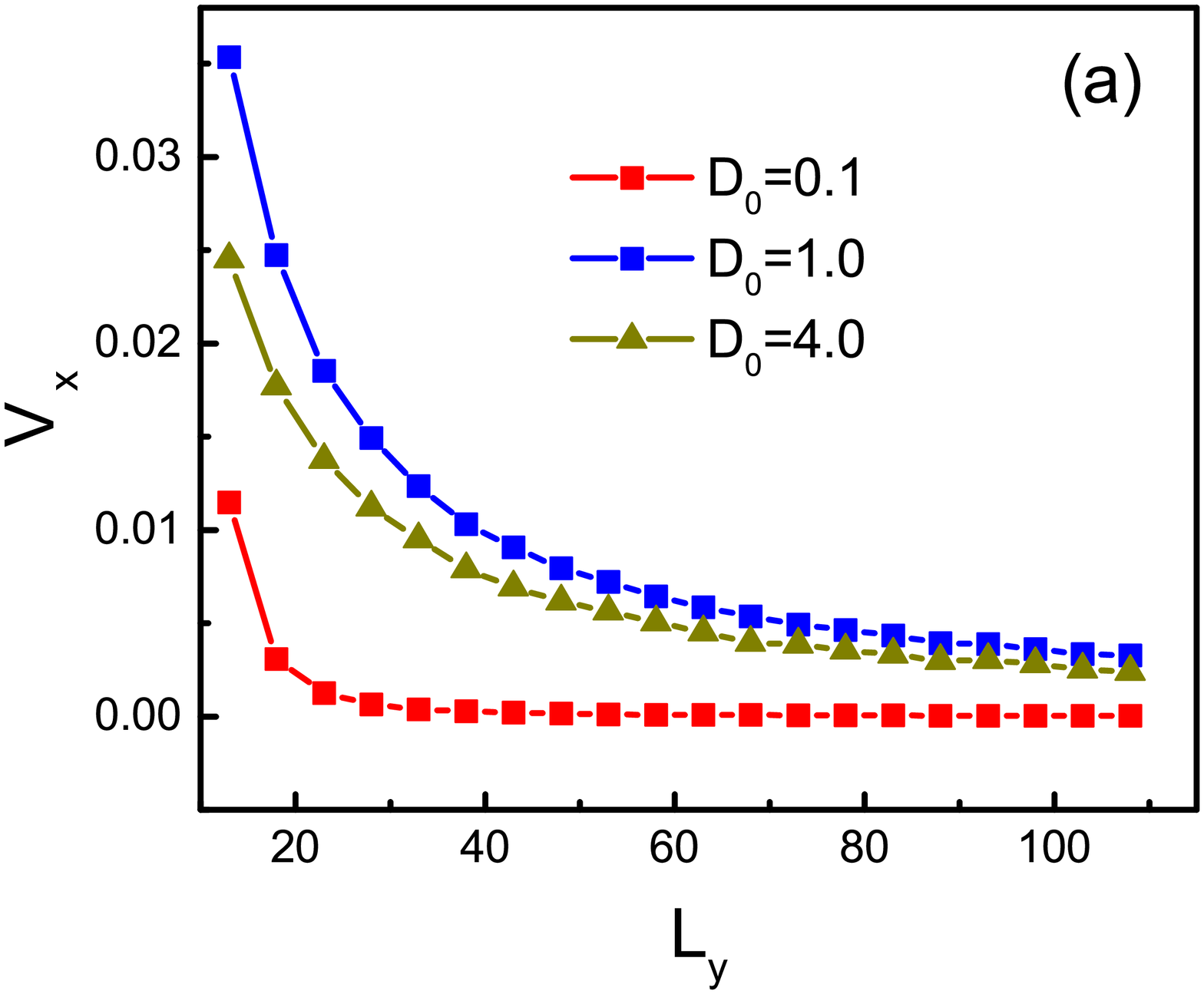}
  \includegraphics[width=0.8\columnwidth]{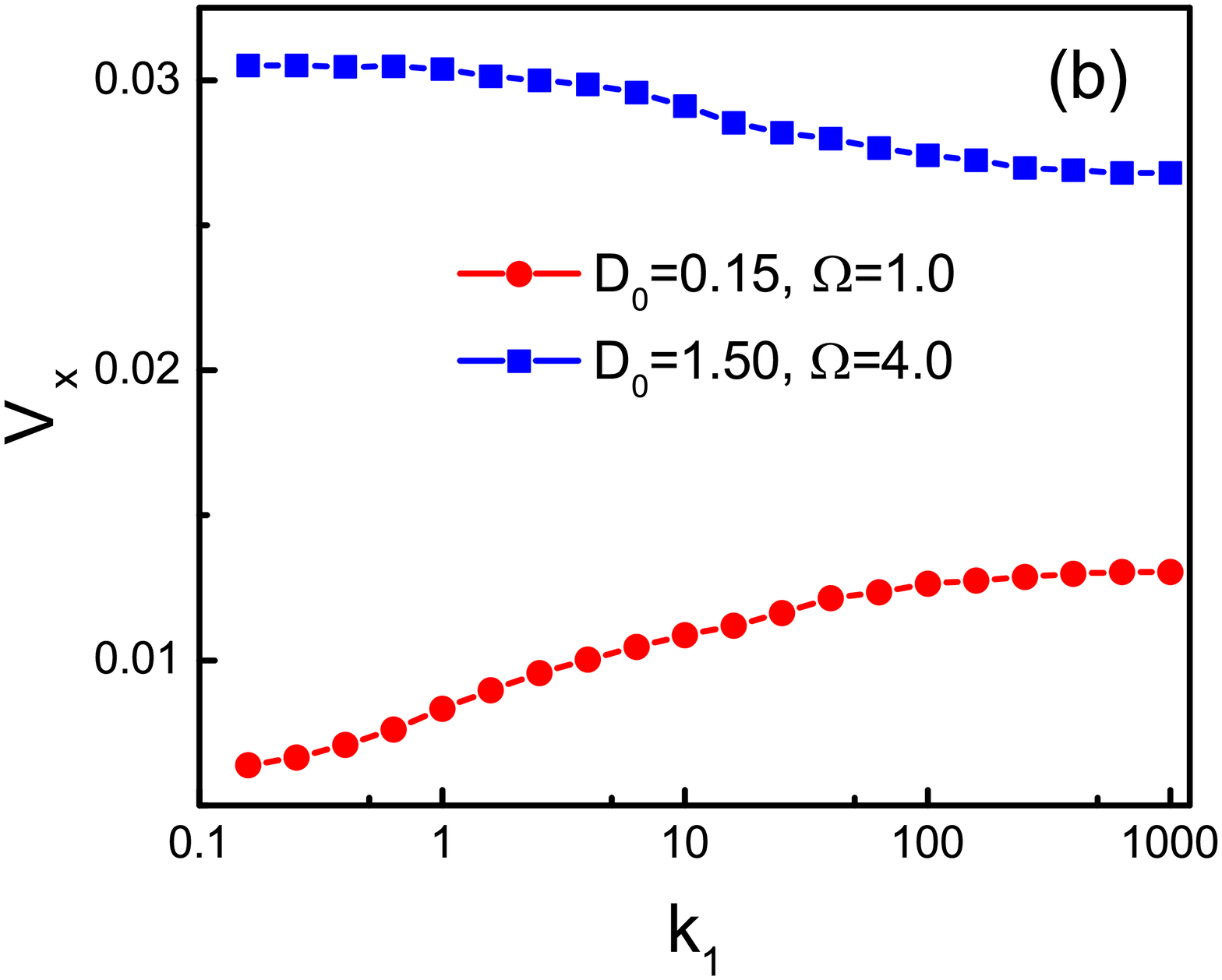}
  \includegraphics[width=0.8\columnwidth]{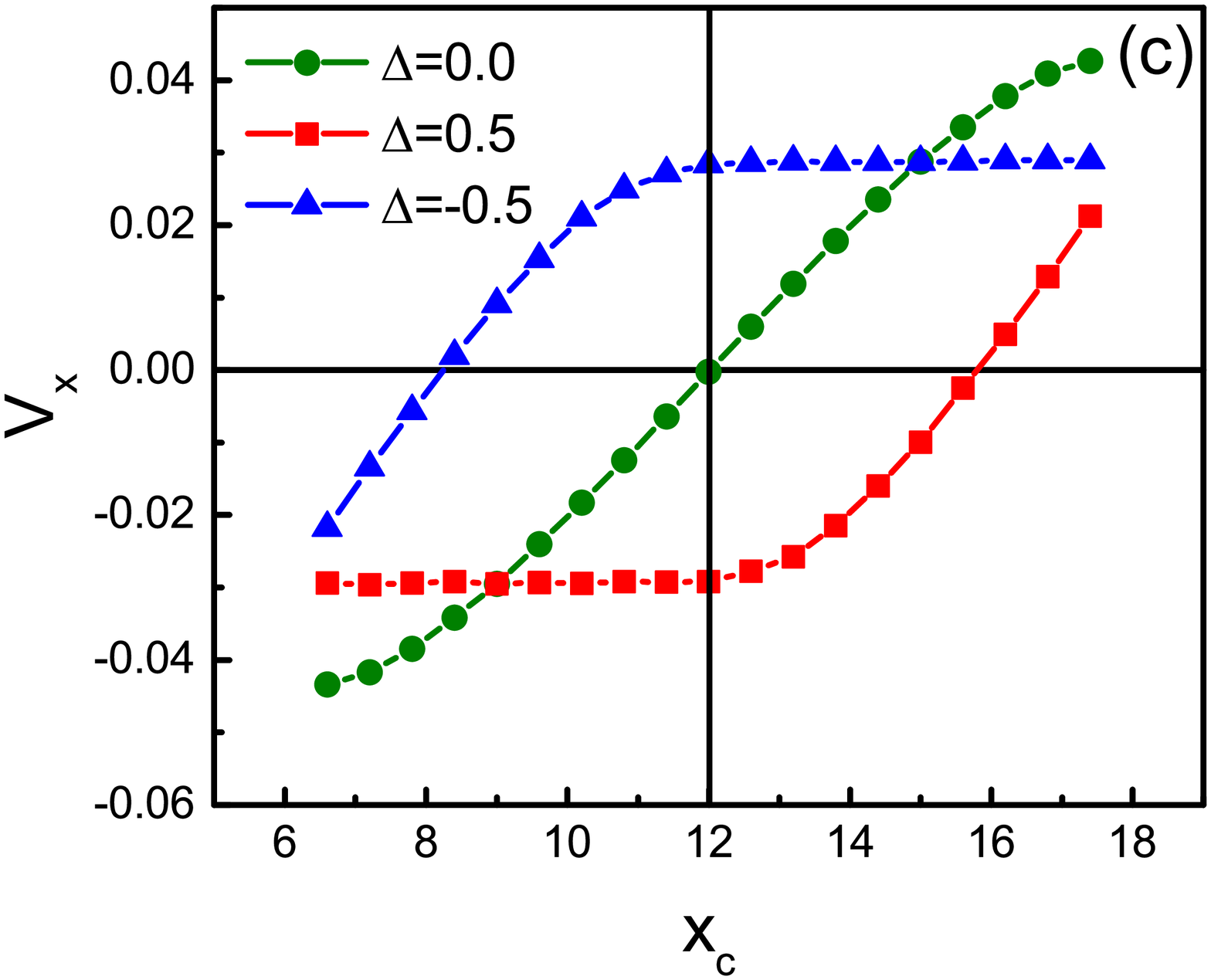}
  \caption{(Color online)(a) Average velocity $V_x$ as a function of the length $L_y$ for different values of $D_0$ at $\Delta=-0.5$ and $\Omega=4.0$. (b) Average velocity $V_x$ as a function of the stiffness constant $k_1$ at $\Delta=-0.5$. (c) Average velocity $V_x$ as a function of $x_c$ at $\Omega=4.0$ for different values of $\Delta$.}
\end{figure}

\indent In order to present more comprehensive information on the rectification, we study the effects of the parameters $L_y$, $k_1$, and $x_c$ on  the rectification for different cases. The dependence of the average velocity $V_x$ on the parameter $L_y$ is shown in Fig. 8(a). In the present system, the length of each paddle is $6$ and $L_y$ must be larger than $12$.  On increasing $L_y$ from $12$, the average velocity $V_x$ decreases monotonically, and finally tends to zero for large $L_y$. When $L_y\rightarrow \infty$, most of particles do not interact with the paddles and particle-paddle interaction becomes insignificant, thus the nonequilibrium driving can be neglected. Therefore, the ratchet effect disappears and average velocity tends to zero.

\indent The role of inter-particle interaction on rectification is shown in Fig. 8(b). The repulsive interaction between particles can (A)disperse Brownian particles (facilitating the directed transport) or (B)reduce the mobility of particles(blocking the ratchet transport). When particles can easily pass across the barrier ($D_0=1.50$ and $\Omega=4.0$),  the factor A dominates the transport, thus the average velocity $V_x$ increases with $k_1$. When particles cannot easily pass across the barrier ($D_0=0.15$ and $\Omega=1.0$), the factor B determines the transport, so the average velocity $V_x$ decreases with increase of $k_1$. Due to short range of the steric repulsive force, on increasing $k_1$ from large value (e. g. 100), the average velocity does not change.

\indent In the above discussion, the wheel is arrayed at the center of the channel ($x_c=12.0$), the position of the wheel in one substrate cell is symmetric. Figure 8(c) describes the dependence of rectification on the position of the wheel for different values of $\Delta$. Because the whole wheel must be in one substrate cell, the position $x_c$ of the wheel must be greater than $6$ and less than $18$.  When the potential is symmetric ($\Delta=0$), it is found that $V_x$ is positive for $x_c>12$, zero at $x_c=12$, and negative for $x_c<12$.  When $x_c=12$, both the potential and the position of the wheel are symmetric, no rectification occurs.  When $x_c>12$, particles can easily pass across the barrier from the right, thus $V_x$ is positive. Similarly, particles easily move to the left ($V_x<0$) when $x_c<12$.  When both the potential and the position of the wheel are asymmetric, the competition between these two asymmetries determined the direction of the transport. For example, when $\Delta=-0.5$, on increasing $x_c$ from $6$ to $18$, the average velocity $V_x$ can change its direction. Therefore, as for the asymmetric potential, varying the position of the wheel is another way of inducing a net current.

\subsection{Transverse ($y$-direction) rectification}
\begin{figure}[htpb]
\vspace{1cm}
  \label{fig3}\includegraphics[width=0.8\columnwidth]{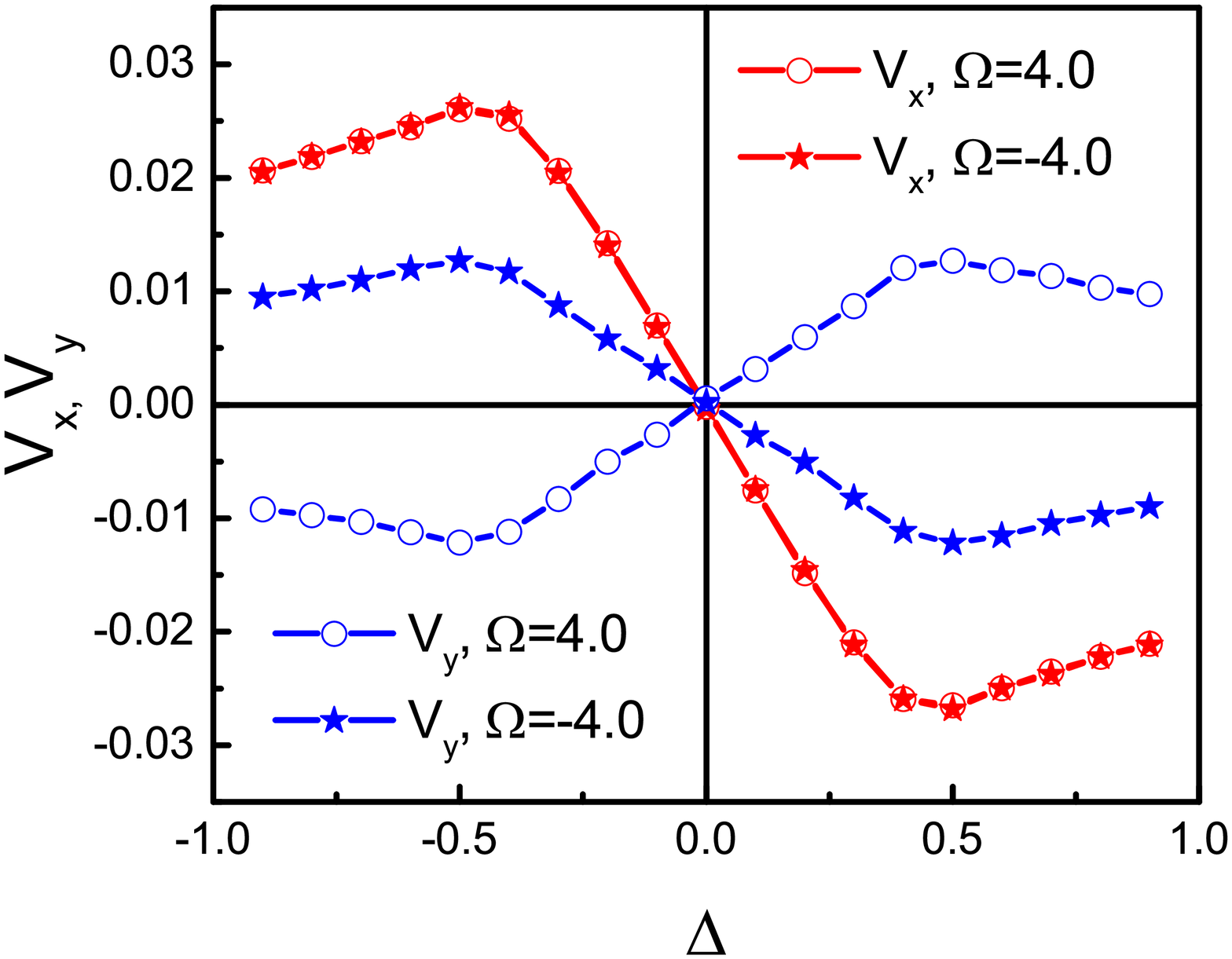}
  \caption{(Color online)Average velocity $V_x$ ($V_y$) as a function of the asymmetric parameter $\Delta$ for both clockwise ($\Omega<0$) and counterclockwise ($\Omega>0$) cases at $D_0=1.0$.}
\end{figure}

\indent In the previous section, particles are confined in the $y$-direction. In order to consider the $y$-direction transport, periodic boundary condition must be imposed in the $y$-direction. Figure 9 shows the average velocity $V_x$ ( or $V_y$) as a function of the asymmetric parameter $\Delta$ for both clockwise ($\Omega<0$) and counterclockwise ($\Omega>0$) cases. Interestingly, the directed transport occurs in the $y$-direction, although the asymmetric potential is only imposed in the $x$-direction. It is found that the sign of $V_y$ is determined by the sign of $\Delta\Omega$, $V_y$ is positive for $\Delta\Omega>0$, zero at $\Delta\Omega=0$, and negative for $\Delta\Omega<0$.  This can be explained as follows. We consider the case of $\Omega>0$ and $\Delta<0$, where particles on average perform the circular motion with counterclockwise direction($\Omega>0$) and the substrate force $f_p(x)$ on the left side is larger than that on the right side($\Delta<0$). Due to the left-right asymmetry of the potential, the motion time along the right side is significantly larger than that along the left side, so particles on average move to the downward side ($V_y<0$). In addition, we find that $V_x$ for the clockwise ($\Omega<0$)case is exactly the same as that for the counterclockwise ($\Omega>0$)case, which indicates that the rotational direction of the wheel does not affect the $x$-direction transport.

\indent Finally, we discuss the possibility of realizing our model in experimental setups. Consider a system of colloidal particles (diameter about $1\mu m$, suspended in an aqueous solution) moving in a two-dimensional channel at room temperature. Two metallic films are deposited on a glass substrate in a periodic but asymmetric fashion, so that applying an ac electric field through electrodes creates an asymmetric potential\cite{Rousslet}. Alternatively, to avoid electrolytic effects, the potential is created optically by strongly focusing an infrared laser beam to form an optical tweezer\cite{Faucheux}. The period of the potential is chosen to be $L_x\approx 24\mu m$ and the width of the channel is $L_y\approx 16 \mu m$. The height of the potential $U_0$ is  $2 k_B T$, where $k_B$ is the Boltzmann constant and $T$ is the temperature.  An uncharged wheel with three paddles (the length is about $6 \mu m$) is arrayed at the center of the channel, which is powered by the external motor. The repulsive hard-core interaction is caused by the enormous increase in energy when two particles overlap. The driving force was transferred from the rotating wheel to colloidal particles, which can break thermodynamical equilibrium and induce the directed transport. Colloidal particles are imaged by a high speed camera, from which the average velocity and effective diffusion coefficient can be calculated. In the experimental setup, we can conveniently control the angular speed and the potential.

\section {Concluding remarks}
\indent In summary, we numerically studied  diffusion and rectification of Brownian particles in the two-dimensional channel, where Brownian particles are driven by the rotating wheel. It is found that the driving force can be transferred from the rotating wheel to Brownian particles, which can break thermodynamical equilibrium and induce the directed transport in the asymmetric potential. (1)When the wheel is located at the center of the channel($x_c=12$), the direction of transport along the potential is completely determined by the asymmetry of the potential. The average velocity $V_x$ is positive for $\Delta<0$, zero at $\Delta=0$, and negative for $\Delta>0$. The rotational direction of the wheel does not affect the $x$-direction transport. The average velocity is a peaked function of $D_0$ (or $\Omega$) and the position of the peak shifts to the large $D_0$ (or $\Omega$) when $\Omega$ (or $D_0$) increases. Interestingly, the giant acceleration of diffusion is observed by suitably adjusting the values of $D_0$ and $\Omega$.  The parameters ($\Omega$ and $D_0$) corresponding to the maximum effective diffusion coefficient are not the same as the parameters at which average velocity is maximum. (2)When the position of the wheel is asymmetric in the channel ($x_c\neq 12$), particles can be rectified even the potential is symmetric ($\Delta=0$). It is found that when the potential is symmetric $V_x$ is positive for $x_c>12$, zero at $x_c=12$, and negative for $x_c<12$, which indicates that varying the position of the wheel is another way of inducing a net current. (3)When the periodic boundary condition was imposed in the $y$-direction. Remarkably, the directed transport occurs in the $y$-direction, although the asymmetric potential is only imposed in the $x$-direction. The direction of the transversal transport is determined by the sign of $\Delta\Omega$.

\section*{Acknowledgements}
\indent This work was supported in part by the National Natural Science Foundation of China (Grants No. 11575064 and No. 11175067), the GDUPS (2016), and the Natural Science Foundation of Guangdong Province (Grant No. 2014A030313426).

\end{document}